\documentclass[]{aa} 
\usepackage{graphicx}   
\usepackage{natbib}   
\usepackage{color}   
\bibpunct{(}{)}{;}{a}{}{,} 

\usepackage{epsfig}

\begin{document}
\title{Long-term observations of Uranus and Neptune at 90\,GHz
  with the IRAM 30m telescope} \subtitle{(1985 -- 2005)}

\author{C.~Kramer\inst{1,2}, R.~Moreno\inst{3,2}, and A.~Greve\inst{4}}
\offprints{C.~Kramer}
\institute{I. Physikalisches Institut,
           Universit\"at zu K\"oln,
           Z\"ulpicher Strasse 77,
           D-50937 K\"oln, Germany
      \and IRAM,
           Nucleo Central, Avda. Divina Pastora 7,
           E-18012 Granada, Spain
      \and LESIA (LAM - bat.18), 5 Place Jules Janssen, 92195 Meudon, France
      \and IRAM,
           300 Rue de la Piscine,
           Domaine Universitaire,
           F-38406 St. Martin d`H\`eres, 
           France
}
\date{received date; accepted date: 23/01/2008\\
}

\abstract
{The planets Uranus and Neptune with small apparent diameters
  are primary calibration standards.}
{We investigate their variability at $\sim90\,$GHz using archived data
  taken at the IRAM 30m telescope during the 20\,years period 1985 to
  2005.}
{We calibrate the planetary observations against non-variable
  secondary standards (NGC\,7027, NGC\,7538, W3OH, K3-50A) observed
  almost simultaneously.  }
{Between 1985 and 2005, the viewing angle of Uranus changed from
  south-pole to equatorial. We find that the disk brightness
  temperature declines by almost 10\% ($\sim2\sigma$) over this time
  span indicating that the south-pole region is significantly brighter
  than average. Our finding is consistent with recent long-term radio
  observations at 8.6\,GHz by Klein \& Hofstadter (2006).  Both data
  sets do moreover show a rapid decrease of the Uranus brightness
  temperature during the year 1993, indicating a temporal, planetary
  scale change.  We do not find indications for a variation of
  Neptune's brightness temperature at the 8\% level.}
{If Uranus is to be used as calibration source, and if accuracies
  better than 10\% are required, the Uranus sub-earth point latitude
  needs to be taken into account. }


\titlerunning{Long-term 90\,GHz observations of Uranus and Neptune}  
\maketitle

\section{Introduction}

At mm-wavelengths, the planets Uranus and Neptune with small
apparent diameter are frequently used for calibration of astronomical
sources and telescope parameters. In this context it is often tacitly
assumed that they are constant radiators, at least at short term
intervals.  Observations over a decade or a longer period, preferably
with the same telescope and receivers and traceable modifications, may
reveal a long--term variability of Uranus and Neptune.

For a long time, Uranus and Neptune have been separated by a distance
of $\sim$\,22$^{\rm o}$ or less in the sky and thus allow precise
comparative measurements.  \citet{griffin1993} observed Uranus and
Neptune in 1990 and 1992 to derive their brightness temperature in the
millimeter and submillimeter regime.  However, recent microwave as
well as observations at visible and near-infrared wavelengths indicate
that Uranus is variable on time scales of several years \citep[see
e.g.][KH06 in the following]{kh06}.  Due to Uranus' large obliquity of
$82^\circ$, measurements from the earth alternate between observations
of the poles and the equator, i.e. the sub-earth point (SEP) latitude
varies.  KH06 argue that the variations they observed at 3.5\,cm
(8.6\,GHz) are partly caused by the geometrical effect, but partly
also by temporal variations deep inside the Uranus troposphere.
However, this variability has not yet been studied at millimeter
wavelengths.


We investigate whether these long-term variations described above are
noticeable in 20 years pointing observations \footnote{This collection
  of pointing measurements provided also the basis for several
  compilations of quasar flux densities published by
  \citet{steppe1988,steppe1992,steppe1993} and \citet{reuter1997}.
  Unfortunately, most of the data of 1988 and 1989 are lost, for
  unknown reason.}  made with the IRAM 30m telescope at 90\,GHz. By
relating the observations of the planets to nearly simultaneous
observations of the constant secondary calibrators NGC\,7027,
NGC\,7538, W3OH and K3--50A, the measurements of the planets are free
from changes of the telescope performance (reflector adjustments,
receiver upgrades, etc.) but contain, on the other hand, the errors of
the secondary measurements.  The secondaries are small relative to the
half power beam width (HPBW) of $27''$ at 90\,GHz.

\section{Observations, selection criteria}

We use heterodyne observations (0.5 and 1\,GHz bandwidth) made between
1985 and 2005 with the IRAM 30m telescope (Pico Veleta, Spain) at
86\,--\,90\,GHz (3.4\,mm). Most of the data were obtained during
pointing measurements, or extended pointing sessions during the
earlier years \citep[see][]{greve1996}. The data are scans across the
planets, and other sources, including the fitted Gauss\-ian profiles,
their halfwidths, their pointing offsets, and their peak antenna
temperature $T_{\rm A}^{*}$, separately determined for the azimuth ({\tt
  Az}) and elevation ({\tt El}) directions. An archived measurement is
accepted for this analysis if the {\tt Az} and {\tt El} pointing
offsets do not exceed 3 -- 5$''$, i.e. being small compared to the
HPBW, and if the {\tt Az} and {\tt El} full width at half maximum
(FWHM) are within 3 -- 5$''$ of the source convolved value. Under
these conditions the antenna temperatures obtained from the {\tt Az}
and {\tt El} scans agree within $\sim$ 10\,$\%$, and their average
value is used.

The hot-cold-sky calibration method used at the 30m telescope corrects
for atmospheric attenuation and gives the antenna temperature $T_{\rm
  A}^{*}$ [K] of the beam convolved source. 

The aperture efficiency $\epsilon_{\rm ap}$ and forward efficiency
$F_{\rm eff}$ \citep[see][]{downes1989,greve1998b} are regularly
determined ($F_{\rm eff}$ to within $\pm\,5\,\%$, $\epsilon_{\rm ap}$
to within $\pm\,10\,\%$) and the flux density of a point source per
beam is $S_{\rm b} = 2\,({\rm k}/A)\,T_{\rm A}^*\,F_{\rm
  eff}/{\epsilon}_{\rm ap} = 3.904\,T_{\rm A}^*\,F_{\rm
  eff}/\epsilon_{\rm ap}$ [Jy] (with $A$ the geometrical area of the
30m reflector, and k the Boltzmann constant).
It is impossible to recover the actual value $F_{\rm
  eff}/\epsilon_{\rm ap}$ for a certain day in order to derive the
flux density $S_{\rm b}$ from the archived antenna temperature $T_{\rm
  A}^*$. However, from simultaneous observations of the constant
sources NGC\,7027, NGC\,7538, W3OH and K3--50A \citep[see for
instance][]{steppe1993,reuter1998,sandell1994}, with flux densities
given in Table 1, we derived the gain $S_{\rm b,sec}/T_{\rm A,sec}^* =
3.904\,F_{\rm eff}/\epsilon_{\rm ap}$ which we applied to the
measurements of the planets $S_{\rm b,pla} = T_{\rm A,pla}^*/T_{\rm
  A,sec}^*\times S_{\rm b,sec}$.  The derived flux densities of the
planets $S_{\rm b,pla}$ contain also the errors of the measurements of
the standard sources.

We assume that any time variance of the four secondary calibrators is
negligible. NGC\,7538, K3-50A, and W3OH are ultra-compact
{\ion{H}{ii}} regions which are stable. Due to the flat spectrum of
the planetary nebula NGC\,7027 \citep{sanchez1998} the reported flux
density variation at 1.4\,GHz of 0.24\,\% per year \citep{perley2006}
is expected to occur, approximately, also at mm-wavelengths. This
amounts to a change of $\sim\,5\,\%$ in 20 years.  We note that we
usually observed several secondaries per day, using their average to
calculate the planetary fluxes, thereby reducing this slight effect.

%
%
One pointing measurement consists of 2 {\tt Az} and 2 {\tt El} scans.
The error of a pointing measurement, used in the following, is the
rms-value of the four scans.
Almost all observations were made in the elevation range
$\sim$\,20$\degr$ -- 70$\degr$. In this range, the gain-elevation
correction at 90\,GHz is less than 2\,\% \citep[cf.  Table 2
in][]{greve1998a}. We did not correct for this minor effect.

This procedure does not allow the derivation of absolute flux
densities of the planets since their flux densities were used to
derive the flux densities of the standards. In the following we
therefore study relative changes of the planet's flux densities and
disk-averaged brightness temperatures.

\begin{figure}
\psfig{figure=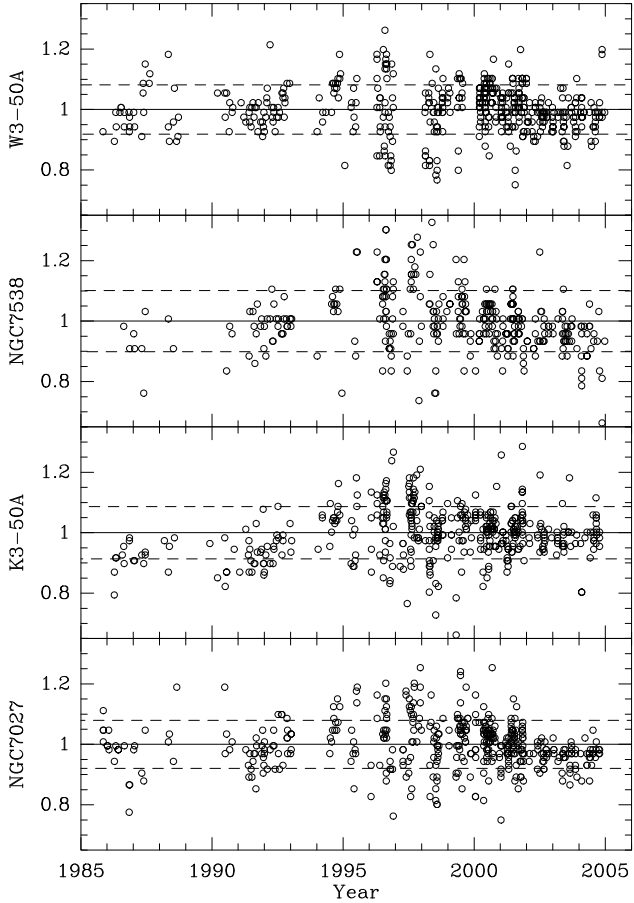,width=8.3cm,angle=0}
\caption[]{ Observations of the secondary calibrators at 90\,GHz.
  Dashed lines show their rms scatter (cf.\,Table\,\ref{tab-secs}). }
\label{fig-secs}
\end{figure}

We show in Fig.\,\ref{fig-secs} the measurements of the four secondary
calibrators over the entire observing period and list their mean
temperatures in Table\,\ref{tab-secs}.  An accuracy of better than
10\% is achieved for all secondaries illustrating the precision of
repeated pointing measurements made by many visiting observers usually
for their own programs.  The observed scatter of antenna temperatures
includes possible changes of the telescope efficiencies from day to
day and over the years. Such changes are calibrated out when studying
relative temperature variations measured on the same day.

From the relation between the observed flux density $S_{\rm b}$, the
Planck function $B$ at temperature $T(\theta,\psi)$,
the beam pattern $P(\theta,\psi)$, and the solid angle $\Omega$ of the
planet subtended at the time of observation

\begin{equation}
S_{\rm b} = {\int_{0}^{\Omega}}{\rm B}[T({\theta} - {\theta}',{\psi} - {\psi}')]
P({\theta}',{\psi}')d{\Omega}'
\label{eq-sb}
\end{equation}

we obtain 
when assuming a constant brightness temperature $T(\theta,\psi) =
T_{\rm B}\,\Pi(\theta,\psi)$ across the disk ($\Pi$) of the planet

\begin{equation}
 S_{\rm b} = (2\,{\rm k}/{\lambda}^{2})\, T_{\rm RJ}\,
 {\int_{0}^{\Omega}}{\Pi}({\theta}-{\theta}',{\psi}-{\psi}')P({\theta'},{\psi}')
 d{\Omega}' 
\label{eq-sbrj}
\end{equation}

and with the Rayleigh-Jeans temperature \newline $T_{\rm RJ} = {\rm h}\,\nu/{\rm
  k}\,(\exp({\rm h}\,\nu/({\rm k}\,T_{\rm B}))-1)^{-1}$ 
\vspace*{-0.5cm}

\begin{eqnarray}
 T_{\rm RJ} & = & ({\lambda}^{2}/2\,{\rm k})\,
 S_{\rm b}\,\left({\int}_{0}^{\Omega}{\Pi}\,P\,d{\Omega}'\right)^{-1} \\
   T_{\rm RJ}           & \equiv & 
        ({\lambda}^{2}/2\,{\rm k})\,S_{\rm b}\left(\Omega\, K \right)^{-1}.
\label{eq-k}
\end{eqnarray}

%
%
The first-order Planck correction at 90\,GHz is \newline $T_{\rm RJ}
\approx T_{\rm B} - {\rm h}\,\nu/(2{\rm k}) = T_{\rm B} - 2.1\,{\rm
  K}$.

In the above equations, we use for the beam pattern $P(\theta,\psi)$ a
Gaussian profile of 27$''$ HPBW \citep{greve1998b}, as it remained
constant throughout the years by using receivers of simi\-lar
illumination taper. Using the IRAM in-house program {\tt planets} we
derive from the flux densities the corresponding disk-averaged
brightness temperature $T_{\rm RJ}$ [K]. This program takes into
account the apparent diameter of the planets $\theta_{\rm s}$ and the
HBPW. In Eq.\,\ref{eq-k}, the correction factor $K$ for a
non-pointlike planetary disk is \citep[cf. Eq.\,12 in][]{baars1973}:

\begin{equation}
K = \frac{1-\exp(-x^2)}{x^2} \equiv \frac{1/f}{x^2} {\rm \,\,with\,\,} 
x = \frac{\theta_{\rm s}}{\theta_{\rm b}} \sqrt{\ln2}
\label{eq-kx}
\end{equation}

where $f$ is called beam dilution factor.

The apparent surface area of Uranus changed by 2.5\% over the 20 years
observing period discussed here. This is due to its oblateness,
coupled with its high obliquity. We take this geometrical effect into
account when discussing possible temporal changes of its disk averaged
temperature in Sec.\,\ref{sec-uranus}.  The variations due to
Neptune's oblateness are much smaller and ignored.

\begin{table}
  \caption[]{Observations  of the secondary calibrators at 90\,GHz for
    the years 1985 to 2005 and the adopted flux densities and
    sizes at this frequency. $N$ is the number of observations.
  }
\begin{center}\small
\begin{tabular}{lrrrcc} 
\hline \hline
Source  & $\langle T_{\rm A}^{*} \rangle$ & rms & $N$ & $S_{\rm b}^{(1)}$ & $\theta_{\rm S}^{(4)}$   \\
        & [K]                             & [$\%$] &  &      [Jy]            & [$''$]      \\
\hline         
W3OH$^{(2,3)}$      & 0.63  &  8.2 & 517 & 3.77 & 14\,$\times$\,10 \\
NGC\,7538           & 0.41  & 10.1 & 351 & 2.37 & 20\,$\times$\,18  \\
K3--50A$^{(3)}$     & 1.06  &  8.7 & 451 & 6.00 & 10\,$\times$\,5    \\
NGC 7027$^{(2)}$    & 0.77  &  7.9 & 455 & 4.58 & $\sim$\,10    \\   
\hline
\end{tabular}
\label{tab-secs}
\end{center}
{\small References: $(1)$ \citet{steppe1993}, for comparison: $(2)$
W3OH:\,4.08\,$\pm$\,0.12\,Jy,
NGC\,7027:\,5.27\,$\pm$\,0.15\,Jy\,\citep{ulich1981}; $(3)$
W3OH:\,3.93\,Jy, K3--50A:\,6.31\,Jy\,\citep[Table\,3
in][]{reuter1998}; $(4)$ approximate dimensions at mm-wavelengths from
\citet{sandell1994}, for comparison: \citet{reuter1998} give at
90\,GHz, W3OH: 6$''$ and K3--50A: 2$''$.}
\end{table} 

\begin{figure}
\psfig{figure=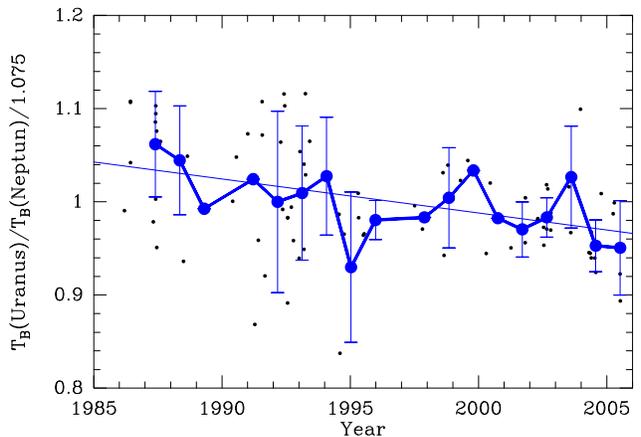,width=8.3cm}
\caption[]{ Ratio of Uranus and Neptune brightness temperatures at
  86-90\,GHz normalized to the average ratio 1.075.  The solid line is
  the result of a linear fit to the 81 observed data (small dots).
  Large dots are the corresponding yearly averages. }
\label{fig-u-n}
\end{figure}

\section{Uranus and Neptune}

Because of their small apparent diameters of $\sim$\,2$''$, Uranus and
Neptune are good calibration standards, at least for the larger and
more sensitive mm--wavelength telescopes. In addition, they were never
far apart in the sky during the past 20 years (distance $\la\,22^{\rm
  o}$) and are therefore ideal for comparative measurements.

\citet{griffin1993} determined Uranus and Neptune brightness
temperatures in the 0.35 to 3.3\,mm region, using Mars as the primary
standard.  From measurements made in 1990 and 1992 they derive
brightness temperatures of Uranus and Neptune at 90\,GHz of $T_{\rm
  B}$(Uranus) = 136\,K and $T_{\rm B}$(Neptune) = 131\,K.
%

Since Uranus and Neptune are close together on the sky, a plot of the
ratio of their brightness temperatures derived from simultaneous
measurements made with the same telescope, may reveal also a relative
change of their brightness temperatures.  Such measurements of Uranus
and Neptune made at frequencies between 86 and 90\,GHz during 1986 and
2005, are shown in Fig.\,\ref{fig-u-n}.  These data were partly
observed without complementary observations of the secondaries.
However, only their relative ratios are discussed in this paragraph.

Calculating the brightness temperature ratio from the ratio of antenna
temperatures, we took into account the Uranus (U) and Neptune (N)
apparent diameters at the time of the observation: $T_{\rm B}({\rm
  U})/T_{\rm B}({\rm N})=T_{\rm A}^*({\rm U})/T_{\rm A}^*({\rm
  N})\times f({\rm U})/f({\rm N})$, with the beam dilution factor
$f(\theta{\rm s},\theta_{\rm b})$ (cf. Eq.\,\ref{eq-kx}). The average
planetary diameter $\theta_{\rm s}$ was calculated from the equatorial
and polar radii (JPL's {\tt HORIZONS} system). In the case of Uranus,
we took into account that the effective polar diameter changes with
viewing geometry (see below).

The ratio is expected to be constant at 1.04 \citep{griffin1993}.  The
average ratio derived from the IRAM observations is 1.07, only
slightly larger.  We are, however, interested in the relative changes.
The rms of the observed ratios is only 6\%. 

A linear least-squares fit to the data \citep{bevington1992} shown in
Figure\,\ref{fig-u-n} results in a slope of $s=-0.4\,10^{-2}$ per
year, i.e. a drop by 8\% between 1985 and 2005. However, the
statistical error of the slope is very large and the
linear-correlation coefficient is small ($r=-0.37$).  In order to
study where this variation may come from, we present in the following
the Uranus and Neptune data independently. These data were observed
near simultaneously with the secondary calibrators to derive their
brightness temperatures.

\begin{figure}
\psfig{figure=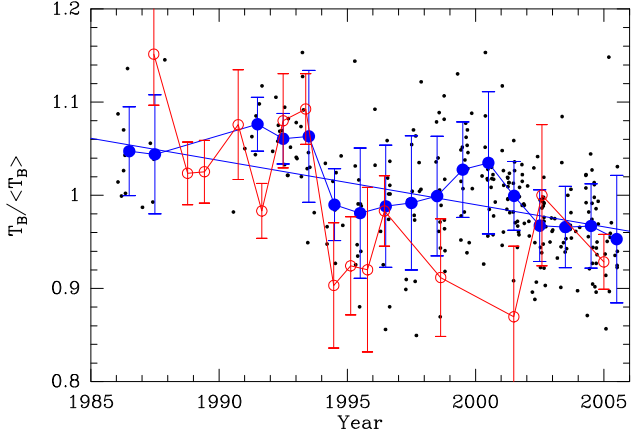,width=8.3cm}
\caption[]{Relative change of Uranus' brightness temperature at
  90\,GHz and at 8.6\,GHz.  The solid line is the result of a linear
  fit to the 286 observed 90\,GHz data (small dots). Their rms is 6\%.
  Large solid dots are the yearly averages (cf.
  Table\,\ref{tab-uranust}).  $\langle T_{\rm B}\rangle$ is the
  average brightness temperature at 90\,GHz data of 134\,K.  Open
  circles show the 8.6\,GHz data of KH06 normalized to $T_{\rm
    lit}=218$\,K which was extracted from Fig.\,3 of KH06 for the
  period shown.  }
\label{fig-uranus-t}
\end{figure}

\begin{figure}
\psfig{figure=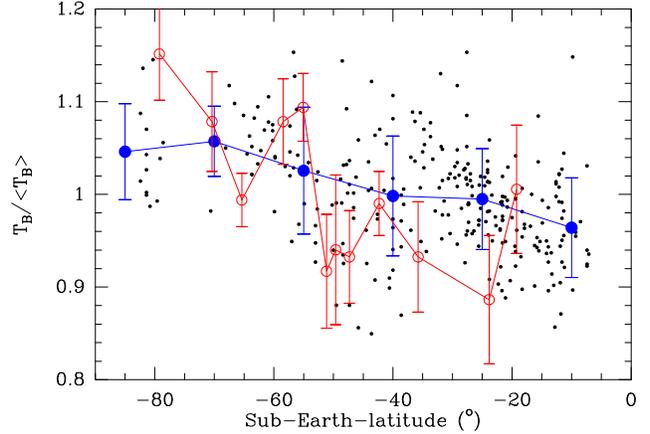,width=8.3cm}
\caption[]{Relative change of Uranus' brightness temperature at
  90\,GHz and at 8.6\,GHz as function of SEP latitude. Small filled
  dots show the individual 90\,GHz measurements, large filled dots
  show the 90\,GHz data binned into intervals of 15$^\circ$.  
%
%
  Errorbars are calculated from the scatter of the individual data.
  Open circles show the 8.6\,GHz data of KH06.  }
\label{fig-uranus-des}
\end{figure}

\subsection{Uranus}
\label{sec-uranus}

Uranus has a 84\,year orbital period of which we have observed
approximately one season (Fig.\,\ref{fig-uranus-t}), between solstice
in the mid 1980\,s and almost equinox which will take place in 2007.
Uranus is slightly oblate, the polar radius of 24973\,km is 2.4\%
smaller than its equatorial radius of 25559\,km
\citep{lindal1987,lindal1992}.  In addition, Uranus has an obliquity
of $82^\circ$.  In 1985, Uranus showed its south pole to the
earth-bound observer, the sub-earth point (SEP) latitude was
$-82^\circ$, the apparent radius was nearly the equatorial radius. It
showed its equator in 2005 when the SEP latitude was $-6^\circ$ (cf.
Fig.\,\ref{fig-uranus-des}).  In summary, Uranus showed different
regions and the disk area changed by $\sim2.5\%$ over the past
20\,years.  The change of disk area has been taken into account in the
following.

The secular change in normalized brightness temperature of Uranus
between 1985 and 2005 at 90\,GHz is shown in Fig.\,\ref{fig-uranus-t}.
These data show an overall scatter (rms) of, again, 6\%.  A linear
least-squares fit to the data results in a slope of $s=-0.48\,10^{-2}$
per year, i.e. a drop by 9\% between 1985 and 2005. Again, the
statistical error of the slope is large and the linear-correlation
coefficient is small ($r=-0.35$). The yearly averages vary between
$+8\%$ in 1991 and $-5\%$ in 2005 (cf.  also Table\,3).  Note that
this drop is consistent within the errors with the change found when
studying the ratio of Uranus and Neptune brightness temperatures
(Fig.\,\ref{fig-u-n}), however with much fewer data. This indicates
that Neptune temperatures are constant.

Figure\,\ref{fig-uranus-t} also shows the results of \citet[][]{kh06}
who studied Uranus at 8.6\,GHz (3.5\,cm) \citep[see
also][]{hofstadter_butler2003} over the last 36\,years and find
significant variability. In late 1993, they find a strong temperature
decrease, indicating a rapid, planetary-scale change.  And indeed, the
90\,GHz data also show a temperature-drop at this time.  Here, the
decrease is by $\sim7\%$.

A plot of the 90\,GHz temperature variation against SEP latitude
(Fig.\,\ref{fig-uranus-des}) shows an almost monotononic decrease of
almost 10\%, from about $+5\%$ at SEP latitudes of about $-85^\circ$
to $-4\%$ at SEP latitudes of about $-10^\circ$, indicating that
Uranus brightness temperature depends on the viewing angle. This
decrease is detected at the $\sim2\sigma$-level as the binned data
have an average rms of $5.4\%$ (Table\,\ref{tab-uranusdes}).
\footnote{Using the equatorial radius for the calculation of the disk
  area at all times, i.e.  not taking into account the change of disk
  area with viewing angle, would lead to a slightly steeper slope and
  a decrease of $-7\%$ at SEP latitudes of $-10^\circ$. }  On average,
the 8.6\,GHz data of KH06 show a similar dependence with year and SEP
latitude, however with larger scatter.

\begin{figure}
\psfig{figure=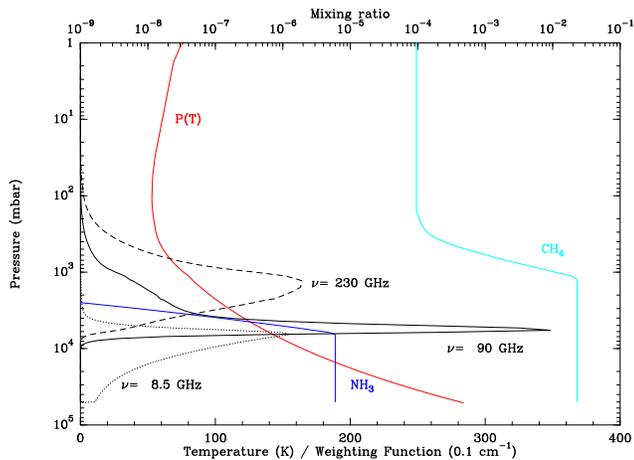,angle=-90,width=8.3cm} 
\caption[]{ Uranus thermal profile $P(T)$ \citep{lindal1987} and
  vertical distribution of NH$_3$ and CH$_4$ mixing ratios. The values
  of the NH$_3$ and CH$_4$ mixing ratios below their condensation
  levels are equal to 6$\times$10$^{-6}$ and 2.3$\times$10$^{-2}$,
  respectively.  These have been used as input to a radiative transfer
  model to calculate the weighting functions of Uranus' continuum.
  The model had previously been used for Jupiter \citep{moreno2001}.
  The weighting function at 8.5, 90 and 230 GHz are shown in black.}
\label{fig-uranus-profile}
\end{figure}

Figure\,\ref{fig-uranus-profile} shows that higher frequencies probe
successively higher layers of the Uranus atmosphere.  The continuum at
8.5\,GHz stems from the high pressure zone, i.e. 11.0 to 5.7\,bar at
low altitudes, where the opacity is sensitive to the NH$_3$ vertical
distribution and the thermal structure.  In contrast, 90\,GHz emission
probes the region of pressures between 6.1 and 4.7\,bar, higher in the
atmosphere, where the opacity is also a function of NH$_3$, but much
more sensitive to the thermal structure.  At still higher frequencies,
the 230\,GHz continuum emission arises from still higher atmospheric
levels (3.5-0.8 bar), near the tropopause. The continuum at 230\,GHz
is mainly sensitive to the thermal structure, and only slightly to the
CH$_4$ vertical distribution, but not to NH$_3$.

It is important to note that between 8.5 and 90\,GHz, the difference
of the opacity sensitivity to NH$_3$ and to the thermal structure, may
explain the differences of the relative changes of Uranus' brightness
temperature between 8.5 and 90\,GHz shown in
Figures\,\ref{fig-uranus-t} and \ref{fig-uranus-des}.

\begin{table}
  \caption[]{Yearly averages and rms of Uranus'
    brightness temperature (cf. Fig\,\ref{fig-uranus-t}). $T_{\rm B}$ is
    the yearly average, $\langle T_{\rm B}\rangle$ is the overall average
    of 134\,K. $N$ is the number of observations. }
\label{tab-uranust}
\begin{center}\small
\begin{tabular}{lcrr} 
\hline \hline
 Year  & $T_{\rm B}/\langle T_{\rm B}\rangle$ & rms & $N$ \\
       &                                      & [\%] \\
\hline 
 1986 &   1.05 &   4.6 &     8 \\
 1987 &   1.04 &   6.1 &     5 \\
 1991 &   1.08 &   2.7 &     6 \\
 1992 &   1.06 &   2.6 &    12 \\
 1993 &   1.06 &   6.7 &    14 \\
 1994 &   0.99 &   3.9 &    10 \\
 1995 &   0.98 &   7.1 &    11 \\
 1996 &   0.99 &   6.6 &    14 \\
 1997 &   0.99 &   7.3 &    16 \\
 1998 &   1.00 &   6.4 &    18 \\
 1999 &   1.03 &   5.0 &    19 \\
 2000 &   1.03 &   7.4 &    21 \\
 2001 &   1.00 &   3.7 &    38 \\
 2002 &   0.97 &   4.0 &    32 \\
 2003 &   0.97 &   4.5 &    19 \\
 2004 &   0.97 &   4.7 &    28 \\
 2005 &   0.95 &   7.2 &    14 \\
\hline
\end{tabular}
\end{center}
\end{table} 

\begin{table}
\caption[]{Normalized Uranus brightness temperatures as function of
SEP latitudes binned into intervals of 15$\degr$ width (see also
Fig.\,\ref{fig-uranus-des}). $N$ is the number of observations.  }
\label{tab-uranusdes}
\begin{center}\small
\begin{tabular}{ccccr} 
\hline \hline
SEP  & $\langle$\,Year\,$\rangle$ & $T_{\rm B}/\langle T_{\rm B}\rangle$ & 
rms &  $N$ \\
Latitude    &      &  [K]              & [$\%$] &    \\
\hline
 -85.0 & 1986.70 &  1.05 &   5.0 &     13 \\
 -70.0 & 1991.92 &  1.06 &   3.6 &     13 \\
 -55.0 & 1994.16 &  1.03 &   6.7 &     44 \\
 -40.0 & 1998.17 &  1.00 &   6.5 &     57 \\
 -25.0 & 2001.67 &  0.99 &   5.5 &    108 \\
 -10.0 & 2004.55 &  0.96 &   5.6 &     51 \\
\hline
\end{tabular}
\end{center}
\end{table} 
\subsection{Neptune}

\begin{figure}
\psfig{figure=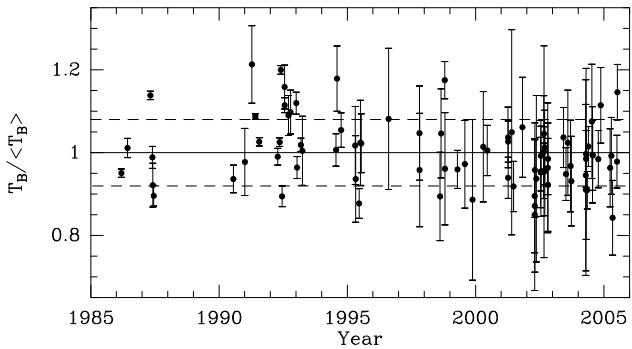,width=8.3cm}
\caption[]{Normalized Neptune brightness temperature. Small dots and
  errorbars show the individual observations and their rms errors.
  Dashed lines show the 8\% (10\,K) rms scatter of all 86
  observations. }
%
%
\label{fig-neptun}
\end{figure}

Figure\,\ref{fig-neptun} shows the brightness temperatures of Neptune,
normalized to the average temperature of $\langle T_{\rm B}\rangle$ of
128.3\,K.  A linear least-squares fit indicates a very small drop of
only 2\% between 1985 and 2005.  In addition, the statistical
uncertainty of the fitted slope is large, 80\%. We conclude that
Neptune temperatures are constant within the 8\% observational error.

%

%
%

\section{Summary}

By careful selection of data from the archived pointing measurements
made at the IRAM 30m telescope during the period 1985 to 2005, we
have been able to study the long-term behaviour of Uranus and
Neptune at 90\,GHz. The large number of selected observations allowed a
statistically meaningful analysis.

\begin{enumerate}

\item We obtained 286 observations of Uranus. The scatter of the
  derived normalized brightness temperatures is 6\%. 

  We find a systematic variation of the Uranus brightness temperature
  of $\sim10$\% with sub-earth point (SEP) latitude when the
  orientation of Uranus changes from south-pole view to equator view.
  This effect is detected at the $\sim2\sigma$ level, the mean scatter
  of the data averaged in $15^\circ$ bins is 5\%. A similar change is
  indicated by the 8.6\,GHz observations of KH06.  The year 1993 shows
  a rapid decrease of brightness temperatures at 90\,GHz of $\sim7\%$,
  which corresponds to a similar decrease seen by KH06, indicating a
  temporal change of a large fraction of the Uranus atmosphere.  More
  observations at other frequencies and improved models of the Uranus
  atmosphere, taking into account its latitude structure, are needed
  to better determine the variations with SEP.  If Uranus is to be
  used as calibration source at 90\,GHz, the latitude dependence of
  its brightness temperature needs to be taken into account.

\item For Neptune, we only obtained 71 observations tied to secondary
  calibrators and 81 simultaneous observations with Uranus. Both show
  that its brightness temperature stays constant at a level of better
  than 8\%.

\end{enumerate}

\begin{acknowledgements} 
  
  We would like to thank our referee Mark Hofstadter for helpful
  comments.  During the 20 years 1985 - 2005, many astronomers and
  operators of the 30m telescope contributed to the pointing
  measurements. We thank the many colleagues for their contribution
  which, unnoticed by most of them, allowed the collection of these
  data.
%
  The computer division, J.\,Pe\~{n}alver (IRAM, Spain), and
  M.\,Bremer (IRAM, Grenoble) helped efficiently in the retrieval and
  transfer of the data.  M. Ruiz (IRAM, Spain) provided the Linux
  version of the program {\tt planets}. 

\end{acknowledgements}

\bibliographystyle{aa}
\bibliography{aamnem99,planets90ghz} 

\end{document}